\documentclass[12pt]{article}
\title{
Chiral symmetry breaking and confinement in the
heavy--light $q\bar q$ system in QCD }
 \author{Yu.A.Simonov\\ Institute of Theoretical
and Experimental Physics\\ 117218, Moscow, B.Cheremushkinskaya 25,
Russia}
\newcommand{\be}{\begin{equation}}
\newcommand{\ee}{\end{equation}}
\def\la{\mathrel{\mathpalette\fun <}}

\def\fun#1#2{\lower3.6pt\vbox{\baselineskip0pt\lineskip.9pt
\ialign{$\mathsurround=0pt#1\hfil
##\hfil$\crcr#2\crcr\sim\crcr}}}

\begin{document}
\maketitle

\begin{abstract}

The effective quark Lagrangian is derived from the QCD action by
means of averaging over vacuum gluonic fields.
In the limit of large $N_c$ and keeping for simplicity the lowest order
gluonic correlator one obtains  nonlinear equations for
the quark propagator, exhibiting scalar confinement.  Connection to the quark
zero mode density is discussed.  \end{abstract}

\section{Introduction}

The chiral symmetry breaking (CSB) in QCD together with confinement
are two most important nonperturbative phenomena, which define the
structure of vacuum and hadrons.
  The simplest situation where both these phenomena define the
 dynamics is the system of one light quark and a very  heavy
antiquark, to be considered as a static source.

In the case of QED the relativistic system of a light and heavy
charges was considered among other things in [1].

The Feynman--Schwinger representation [2] studied in [1] allows
to make the infinite mass limit and obtain the effective Dirac
equation for the light charged particle (this is in contrast to
Bethe--Salpeter equation, where the one--body limit is not
recovered at least in lowest  order approximation for the
kernel).

The attempt to extend these results to QCD [3] meets with serious
difficulties, and one needs the real understanding of dynamics
of confinement and CSB in their interconnection.

While confinement was
explained recently as being produced by specific vacuum
correlators of gluonic fields [4] (of the form of correlators of
effective magnetic monopole currents), the phenomenon of CSB is still
not understood.

 In this paper we
 suggest a systematic method to treat CSB in connection with
 confinement.  We start with the QCD
 Lagrangian and derive from that the effective Lagrangian of light
 quarks, assuming that certain gluon field correlators are nonzero,
which are known to yield confinement (i.e.  linear potential) for
static quarks. It is not clear from the beginning what will happen
for light quarks ( with vanishing mass) and whether they would be
confined at all.

Our main concern in what follows is to keep Lorentz and gauge
invariance. The effective quark interaction is nonlocal and to have
gauge-invariant equation one should consider $q\bar q$ Green's
functions.

As mentioned above the simplest setting for which confinement and
CSB can be studied in the gauge--invariant way, is the problem of a
light quark propagation in the field of the static antiquark. To
simplify matter we start with the Gaussian correlators for gluon
fields and derive the selfconsistent  equations for the light quark
propagator (with a string effectively connecting it to the static
source). We show that CSB   occurs due to the string (linear
confinement), which shows up in the fact that effective interaction
becomes Lorentz scalar.
 As another
evidence of CSB  in [5] the chiral condensate is computed and shown to be
nonzero.
The strongly modified and extended version of this study  in
[5] contains also another  derivation of scalar confinement.

Those results were derived actually for the case of one light
quark (quenching approximation or $N_c\to \infty$ ).
 To treat the case of two and more light quarks the
corresponding term in the effective  Lagrangian may be studied, and
one is naturally led to the equation for the Green's  function of two
(or more)
light quarks.
The study of two and three quark systems can be performed in the framework of
the same formalism and is  now in progress.

 The paper is organized as follows. In the second section the
general form of the effective Lagrangian for light quarks is  given,
and the equation for the Green's function of one light quark in the
field of the static source is derived. In the third section this
equation is studied and  solutions are derived,
which  exhibit scalar confinement and hence CSB.

In the fourth section the connection of the chiral condensate to
quark zero modes is illustrated in the framework of the present
approach.

The conclusion is devoted to a summary of results and the discussion
of prospectives.

\section{Derivation of the effective Lagrangian for the light quark}

To make discussion of confinement and chiral symmetry breaking fully
gauge invariant, we consider the gauge invariant physical amplitude
-- the Green's function $S$ of a light quark of mass $m$ in the
field of a static antiquark placed at the origin. The propagator for
the latter
in positive time  direction
 can be taken as (we consider below Euclidean space-time).
\be
S_{\bar Q}(T) \equiv \frac{1+\gamma_4}{2}P~exp~ig\int^T_0A_4(\vec
r=0,\tau) d\tau
\ee

The Green's function $S_{q\bar Q}$ can be written as an integral
\be
S_{q\bar Q}(x,y)=\frac{1}{N}\int
D{\psi}D{\psi^+}DA
e^{-\int\frac{F_{\mu\nu}^2}{4} d^4x
-\int{\psi^+}(-i\hat \partial -im -\hat A)\psi d^4x}
\psi^+(x) S_{\bar Q} (x,y)\psi(y)
\ee

Since $S_{q\bar Q}$ is gauge invariant, one can choose any convenient gauge
for $A_{\mu}$ and our  choice will be the modified Fock--Schwinger gauge,
 namely for any point $x$  one introduces the curve $C(x,s), s\leq  1$ with
the tangent vector $t_{\mu}(s)$ and coordinate $z_{\mu}(x,s)$.
The gauge condition
(for some general class of the curves
$C(x,s)$ considered below) is
 \be
 A_\mu(z)
t_{\mu}=0.  \ee

We choose the curve $C$ to go from the point $x$ along some path
(to be specified later) to the
4-axis, and then run along it, so that
$A_4(\vec r=0,\tau)=0$.
 In this gauge $G_{\bar Q}$ reduces to
the factor $\frac{1+\gamma_4}{2}$ and one can now consider the integration
over $DA$ in (2) as a statistical averaging process and use the cluster
expansion [6], ($a,b$~-~ color~indices)
$$ \int DA e^{\int\psi^+(x)\hat
A(x)\psi(x)d^4x-\int\frac{1}{4} F^2_{\mu\nu}d^4x}=<exp
\int  \psi^+_a(x)\hat
A_{ab}(x)\psi_b(x) d^4x> $$ \be =exp\{\int
d^4x\psi^+(x)\gamma_{\mu}<A_{\mu}>\psi(x)+ \ee $$+\frac{1}{2}\int
dxdy\psi^+(x)\gamma_{\mu}\psi(x)
\psi^+(y)\gamma_{\nu}\psi(y)<A_{\mu}(x)A_{\nu}(y)>+...\}
$$

We have denoted the higher order cumulant contribution in (4) as
$...$ and shall disregard it
here, for the discussion  of these terms the reader is referred to
[5].
The first term $<A_{\mu}>$ vanishes due to the gauge and Lorentz
invariance of the vacuum, while the second can be expressed through
the field strength correlators $<F_{\mu\nu}(u)F_{\lambda\sigma}(u')>$
in the gauge (3) as follows
\be
A_\mu (x)=\int_C ds\frac{dz_{\alpha}(s,x)}{ds} F_{\alpha\beta} (z)
\frac{dz_{\beta}}{dx_\mu}
 \ee

The representation (5) satisfies condition
(3)
for the class of curves $C$ considered below.
 Using (5) one can rewrite the average $<AA>$ in (4) as
$$
<A^{ab}_{\mu}(z)A^{cd}_{\nu}(w)>=\frac{\delta_{bc}\delta_{ad}}{N_c}
\int^z_C du_\alpha\int^w_Cdv_{\gamma}\times
  $$
  \be
  \times <F_{\alpha\beta}(u)F_{\gamma\delta}(v)>
\frac{du_\beta}{dz_\mu}\frac{dv_\delta}{dw_\nu}
\ee
  where $ab,cd$ are fundamental color indices.
  The gauge--nonivariance of the correlator $<FF>$ in (6) is only
  apparent and one can introduce a factor, equal to unity in the
  gauge (3), which makes the correlator explicitly gauge--invariant,
  namely
  \be
  <F(u)F(v)>=<F(u)\Psi F(v) \Psi^+>
  \ee
  where $\Psi$ is the product of 2 parallel transporters
  \be
  \Psi= \Phi_C(u)\Phi^+_C(v)
  \ee
  and
\be
  \Phi_C(u)=P~exp~ig \int^u_{C} dz_{\mu}A_{\mu}
\ee
   where the path-ordered contour integral $\Phi$ is taken along the
   contour $C$.

   Since in the gauge (3) one has $\Psi=\Psi^+\equiv 1$, we shall
   below omit those factors. For the correlator $<FF>$ one can use
   the parametrization suggested in  [7]
   \be
   g^2<F_{i\mu}(u)F_{i'\mu'}(u')>_{ab}=
   \delta_{ab}(\delta_{ii'}\delta_{\mu\mu'}-
   \delta_{i\mu'}\delta_{i'\mu}) D(u-u')+\Delta^{(1)}
   \ee
   where $\Delta^{(1)}$ is proportional to a full derivative, its
   exact form is given in [7].

   In what follows we shall consider only the term $D$ in (10)
   since it contributes to the string tension, while $\Delta^{(1)}$
   does not. Namely using (10) it was obtained in [7] that the string
   tension $\sigma$ -- the coefficient in the area law of the Wilson
   loop, $<W(C)>=exp (-\sigma area)$ is equal to
   \be
   \sigma = \frac{1}{2} \int^{\infty}_{ -\infty} d^2uD(u)
   \ee

   Thus $\sigma$ characterizes the confinement of static quarks,
   and our goal is to understand how the dynamics of light quarks is
   expressible through $\sigma$ and whether $\sigma$ correlates with
   CSB.

Keeping only the term $D$ in (10) and neglecting higher order
correlators like $<AAA>$, one obtains in (4) the following effective
Lagrangian for the light quark:
$$
{\cal{L}}_{eff}(\psi^+\psi)=\int \psi^+(x)(-i\hat \partial -im )
\psi(x)d^4x +
$$
\be
\frac{1}{2N_c}\int
d^4xd^4y(\psi^+_a(x)\gamma_{\mu}\psi_b(x))(\psi^+_b(y)
\gamma_{\mu'}\psi_a(y))\times
\ee
$$
\times
J_{\mu\mu'}(x,y)
$$
where we have defined
$$
J_{\mu\mu'}(z,w)=\int^z_C du_\alpha\int^w_Cdv_{\gamma}
(\delta_{\alpha\gamma}
\delta_{\beta\delta}-\delta_{\alpha\delta}
\delta_{\beta\gamma})
\frac{du_{\beta}}{dz_{\mu}}\frac{dv_{\delta}}{dw_{\mu'}}\times
$$
\be
\times D(u-v)
\ee

In what follows we disregard the perturbative contributions to
${\cal{L}}_{eff}$, since they have nothing to do with CSB. The mass
$m$  is supposed to be defined at the typical hadronic scale of 1
GeV, and we shall not be interested in its evolution to lower
scales.

From the effective Lagrangian (12)
in the large $N_c$ limit one can easily derive the equation
of Dyson-Schwinger type for the selfenergy part, which we shall
denote by $M$ and  the $q\bar Q$ Green's function $S$.
(We denote by $S$ the form (2) where instead of $S_{\bar Q}$ only the
parallel transporter is retained (cf. Eq.(1))).
The definition and the diagrammatic representation of $M$
  is done in the
same way, as in the NJL model [8], since the structure of the
Lagrangian (12) is similar to that of  NJL however nonlocal.

The main essential difference is the presence of the string ,
connecting the light quark to the static source, this part is
concealed in $J$, Eq.(13) and therefore the selfenergy part $M$ is
actually not the set of the one--particle--irreducible diagrams, but
rather the $\bar q Q$ interaction kernel.

In the configuration space the equations for $M$ and $S$ are readily
obtained from (12) noting that in the  large $N_c$
limit
one has to replace a pair of $\psi,\psi^+$ operators in (12) as
\be
\psi_b(x)\psi^+_b(y)\to<\psi_b(x)\psi^+_b(y)>=N_cS(x,y),
\ee
and finally one obtains
\be
iM(z,w)=J_{\mu\nu}(z,w)\gamma_{\mu}S(z,w)\gamma_{\nu}
\ee
\be
(-i\hat{\partial}_z-im)S(z,w)-i\int M(z,z')S(z', w)d^4z'=
\delta^{(4)}(z-w)
\ee

The system of equations (15-16)
is exact in the large $N_c$ limit, when higher correlators are neglected and
defines unambiguously both the interaction kernel $M$ and the Green's
function $S$.  One should stress at this point again that both $S$ and $M$
are not the one-particle operators but rather two--particle operators, with
the role of the second particle played by the static source. It is due to
this property, that $S$ and $M$ are gauge invariant operators, which is very
important to take confinement into account properly. Had we worked with
one--particle operators, as is the habit in QED and sometimes also in QCD,
then we would immediately loose the gauge invariance and the string, and
hence confinement.

\section{Chiral symmetry breaking by the solutions of equations (15),
(16)}

As was discussed in the Introduction, CSB may reveal itself in the
heavy-light system in two ways: i) as the presence of the scalar part
in the effective mass operator $M$ and ii) as the nonzero quark
condensate $tr S(0,0)\sim <\bar \psi\psi>$.

Both points have been studied in [5] using the relativistic WKB method and
the presence of scalar confinement and nonzero quark condensate was shown
as a consequence of the selfconsistent WKB solution.

In this paper we shall give another argument in favour of the scalar
confinement, using an approximate
 asymptotic form of the solution of (15-16) at large  distances.

The idea which enables us to find this approximate solution is based on the
following simple physical picture. At large distances the string formed
between the quark and the static source becomes heavy due to the length of
the string even for the massless quark. Therefore the situation
is similar to the case of the heavy quark, where the exact solution is known
in the limit of the infinitely large mass, and one can use an expansion in
powers of inverse mass of the string.

To introduce this expansion we first specify the kernel $J_{\mu\nu}$ in (15)
and to this end define the contour $C$ in (5) and the form of the correlator
$D(x)$ in (10).

For the latter it is convenient to choose  the Gaussian form,
\be
D(x) = D_0 exp (-\frac{x^2}{4T_g^2}), D_0=\frac{\sigma}{2\pi T_g^2}
\ee
where the use of equation (11) was made.

To calculate $J_{\mu\nu}(z,w)$ we choose the special case of the gauge (5),
which may be called "the planar gauge": the contour $C(x)$  is formed by the
three straight--line
 pieces: i) first a perpendicular
to some "gauge" plane, passing through the 4-th axis, then ii) a
perpendicular to the 4-th axis, and finally iii) a cut along the 4-th axis to
$x_4 = -\infty$. One can choose the "gauge plane" passing in the middle
between the points $\vec{z}$ and $\vec{w}$ in $J_{\mu\nu}(z_4 -w_4, \vec{z},
\vec{w})$. To simplify calculations let this plane be the 14 plane,
and then omitting for the moment the contribution of the cut i) in
the integral (5) (we shall discuss its contribution below) one
obtains the following representation of the only nonzero component of
$J_{\mu\nu} = J_{44}$
\begin{equation}
J_{44}(z,w) = e^{-\frac{(z_4 -w_4)^2}{4T_g^2}} \sigma \frac{z_1
w_1}{2\pi T_g^2} \int\limits^1_0 \int\limits^1_0 dsdt
e^{-\frac{(z_1 s -w_1 t)^2}{4T_g^2}}
\end{equation}

In the limiting case when 3-vectors $\vec{z}$ and $\vec{w}$ coincide
and lie in the gauge plane, the expression (18) simplifies
\begin{equation}
J_{44}(z,w) = \frac{\sigma|\vec{z}|}{\sqrt{\pi}T_g} exp (-\frac{(z_4
-w_4)^2}{4T_g^2})
\end{equation}

We now turn to equations (15 - 16) and look for the solution in the
form of an expansion
\begin{eqnarray}
S & = & S_0 + S_1 + ... \nonumber \\
\\
M & = & M_0 + M_1 + ... \nonumber
\end{eqnarray}
with
\begin{equation}
M_0 = J_{44} \gamma_4 S_0 \gamma_4
\end{equation}
and
$$
S_0{(z,w)} = \frac{i}{2}\{\theta(h) (1+ \gamma_4) exp
[-\int\limits_0^h \bar{M}(h', \vec{z}, \vec{z}) dh'] \varphi_0^{(1)}
(\vec{z}, \vec{w}) +
$$
\be
 + \theta(-h)(1-\gamma_4)exp[\int\limits_0^h \bar{M}(h',
\vec{z}, \vec{z})dh'] \varphi_0^{(2)} (\vec{z}, \vec{w}) \}
\ee

Here notations are used
\begin{equation}
h \equiv z_4 -w_4 \; , \;\; \varphi_0^{(1)} = \varphi_0^{(2)} =
\delta^{(3)}(\vec{z} -\vec{w})
\end{equation}
and $\bar{M}$ is to be defined from the solution of (15 - 16).

Acting with the operator $-i \frac{\partial}{\partial h} \gamma_4$ on
(22) one has
\begin{equation}
-i\partial_4 \gamma_4 S_0(h, \vec{z}, \vec{w}) = \delta^{(4)}(z-w)+
i\bar{M}(h_4, \vec{z}, \vec{z})S_0(h, \vec{z}, \vec{w})
\end{equation}
From (21) one obtains
\begin{equation}
M_0(h, \vec{z}, \vec{w}) = \frac{\sigma|\vec{z}|}{\sqrt{\pi}T_g}
\delta^{(3)}(\vec{z}-\vec{w})e^{-\frac{h^2}{4T_g^2}} N
\end{equation}
where
\begin{equation}
N =\theta(h)(1+\gamma_4)exp(-\int\limits_0^h \bar{M} dh')+\theta(-h)
(1-\gamma_4) exp(\int\limits_0^h \bar{M}dh')
\end{equation}

To simplify matter further we take the so-called string limit, i.e.
tend
$T_g$ to zero while keeping string tension $\sigma$ constant. In this
case it will be sufficient to consider $\bar{M}$ independent of time $h'$
in (22), and one obtains
\begin{equation}
\int d^4 z' M_0(z,z')S_0(z', w) = \sigma|\vec{z}|S_0 (z,w)
\end{equation}

In derivation of (27) the condition $\sigma|\bar{z}| T_g \ll 1$ was
used, i.e. the string limit is taken before any other asymptotics is
considered.

The space derivatives in (16) can be  shown to give subleading terms, as
compared to (24) and (27), this is exactly the same situation as in the heavy
mass limit considered in [5]. Indeed, inserting (24) and (27) into (16) on
has
\be
\bar M(\vec z, \vec z) =\sigma |\vec z| +m
\ee
At the same time the spacial derivative $\vec \gamma \vec \partial S_0$ does
not contain the large parameter $|\vec z| \to \infty$, and therefore should
be taken into account in the next order of the expansion (20). The appearence
of the derivative $\vec \gamma\vec \partial \delta^{(3)} (\vec z-\vec w)$
signals that the local approximation (23) used for $S_0$ should be in next
orders replaced by the nonlocal form, which was obtained in [5],
\be
M\sim S(x,y) \sim \frac{\sigma}{\pi^2\sqrt{xy}}K_0(\sigma \sqrt{xy}|x-y|)
\delta (1-\cos \theta_{xy})
\ee
where $K_0$  is the McDonald function. The r.h.s. of (29) is the smeared
$\delta$--function at large $|\vec x|\sim |\vec y|$, which should replace (23)
in the next approximation.

One might improve the initial approximation (22) keeping in (16) the local
form of $M$ (25), (since it enters under the integral) but replacing
$\varphi_0^{(1)}, \varphi_0^{(2)} $ in (22) by some nonlocal form.  Instead
of writing explicit formulas, one can
 obtain the simple static form of equation (16) if one integrates both sides
of (16) over all infinite interval of $h=z_4-w_4$. One obtaines
\be
[-i\vec \gamma\vec\partial -i(m+\sigma|\vec z|)]\bar S(\vec z, \vec w)=
\delta^{(3)}(\vec z-\vec w)
\ee
where we have defined
\be
\bar S(\vec z, \vec w)=\int^\infty_{-\infty} dh S(h,\vec z, \vec w)
\ee

Both equations (28) and (30) signify the scalar confinement at large
distances, independently of the value of the quark mass $m$.

This statement agrees with the conclusion of the paper [5], where
another, less direct method was used.

At this point one can compare our results with those in [9] and
[10], where the limit of the heavy quark mass $m$ was used (the
authors of [9] have exploited the form (15) with $S$ replaced by
$S_0$ as the basic equation), and at the same time to elucidate the
dependence of results on the choice of the gauge.

One should stress first that the gauges used in [5], [9] and [10] on
one side and in the present paper are different.

This fact does not play any role if one considers the total effective
action, i.e. the form (12) with addition of all higher order
correlators as in [5], Eq. (163). However the Gaussian approximation
for $L_{eff}$ and $M$, as in (15) depends on the gauge, or (if one
considers the quark trajectory $L$ with the line $C(x)$ attached at
each point of $L$) on the shape of the gauge surface, this dependence
being cancelled by higher order correlators. To minimize the
influence of higher-order correlators one should choose the shape of
the gauge surface as close to the real physical world sheet of the
string as possible, and the latter is known to be the minimal area
surface.

The situation here is parallel to that
of the perturbative QCD series, where keeping a few first terms
 one obtains a dependence on the normalization mass $\mu$, which should
not enter into the physical result. The usual prescription is to
choose $\mu$ in such a way as to minimize the neglected terms of the
perturbative series and the recip\'{e} is to take $\mu$ of the order
of the inverse size of the system.

In accordance with what was said above it is profitable to choose the
gauge surface of minimal area, and our choice of the plane instead of
the surface in [5], [9] and [10] strongly reduces the contribution
of the spacial projections of the gauge surface. The latter enter in
the final result being multiplied by the color-magnetic correlators
$<F_{ij}(z)F_{lm}(w)>$, $i,j,l,m = 1,2,3$. Comparison of (28) with
the results of [5], [9] and [10] shows that (28) contains only the
contribution of the color-electroc correlator which agrees with that
of [5], [9] and [10], while the latter papers contain also the
contribution of magnetic correlators. Hence the color-electric
contribution is surface (and gauge-) independent, while the
color-magnetic one depends on the shape of the gauge and can be
excluded for the best (mininal) choice of the gauge surface (the
above is true only for the zero orbital momentum, otherwise the
minimal surface is close to the helycoid, rather than to the plane,
and the magnetic contribution survives).

\section{Quark zero modes $vs$ field correlators}

We have seen in previous sections that the nonlinear equations (15),
(16) give rise to the phenomenon of CSB, which reveals itself in our
problem in two ways: i) it provides scalar confining interaction for
the light quark ii) there appears a standard chiral condensate
$<\bar \psi \psi>$.

A natural question arises at this point: a folklore understanding of
CSB is that it is due to quasizero global quark modes in the vacuum.
An exact relation [11] exists, which connects chiral condensate to the
density $\nu(\lambda)$ of quasizero modes in the vacuum at
$\lambda\la m$, and in the chiral limit $(m\to 0)$ one has [11]
 \be
 <\bar \psi \psi>=-\frac{\pi\nu(0)}{V_4}
 \label{7.1}
 \ee
 Here $\lambda$ is an eigenvalue of the 4d Euclidean equation for the
 quark in the vacuum field $A_{\mu}$
 \be
 -i\hat D   \psi_n(x) = \lambda_n\psi_n(x)
 \label{7.2}
 \ee
 The density $\nu(\lambda)d\lambda$ is the averaged over all fields
 $\{A_{\mu}\}$ number of the states $\lambda_n$ per interval
 $d\lambda$.

 It is a popular belief that the quasizero modes necessary for CSB
 due to (32) are descendant from the local zero modes on the
 topological charges (instantons or dyons), and their density is
 therefore proportional to the density of instantons (dyons). There
 are the instanton model [12] and  the dyon model [13] of the QCD
 vacuum, which explain CSB in this way.

 Whether these models are realistic or not, is the open question, but
 the Banks-Casher relation (32) holds independently of that,
 and if the method of the present paper proves CSB due to the field
 correlators (even in the Gaussian approximation), one should explain
 the origin of the quasizero modes in (32).

 To do this we consider first the case of Abelian fields. As was
 stressed above in the paper, CSB is due to the correlator $D(x)$,
 and the latter in the Abelian case can be connected to the
 correlator of magnetic monopole currents [7,4]
 \be
 <\tilde j_{\beta}(x) \tilde j_{\delta}(y)> =
 (\frac{\partial}{\partial x_{\alpha}}
 \frac{\partial}{\partial y_{\alpha}}
 \delta_{\beta\delta}-
 \frac{\partial}{\partial x_{\beta}}
 \frac{\partial}{\partial y_{\delta}} ) D(x-y)
 \label{7.3}
 \ee

 In the nonabelian case one can use the Abelian projection method
 (APM) [14], to separate out of the field $A_{\mu}$ and field
 strength $F_{\mu\nu}$ the monopole and photon part, and the part of
 "charged gluons". The latter contributes around 10\% to the
 effective action.  In this case one can connect the monopole current
  obtained by APM with the correlator $D$ as in (34).

 Now  each magnetic monopole with the world line of the length $T$
 produces $T/b$ quasizero modes, where $b$ is the size of the
 monopole [15]. The interval of modes produced by monopoles is
 $\Delta \lambda$, threfore the 3d density of monopoles $n_3$
 multiplied with 3d volume $V_3$ and with $T/b$ gives the total
 number of quasizero modes in the intervals $\Delta\lambda$, i.e.
 $\nu\Delta \lambda = \frac{T}{b} V_3n_3$.

 Taking into account that $V_4=TV_3$, one has a relation $n_3\sim
 \frac{\nu(0)}{V_4}(\Delta \lambda b)$. The last factor $(\Delta
 \lambda b)$ is equal to unity if the quasizero modes of monopoles
 are mixed due to interaction into the whole interval $\Delta
 \lambda \sim 1/b$.  (As will be seen under this assumption one
 obtains the estimate (35) for the chiral condensate.)
  On the other hand one can estimate
 the 3d density of magnetic monopoles,  from
 $<\tilde j_{\beta}(\vec x, x_4)\tilde j_{\beta}(0,x_4)>$ integrating
 over $d^3\vec x$. (The correlator $<\tilde j(x)\tilde j(0)>$
 estimates probability of finding a monopole at the point $x$, if
 there is one at $x=0$.  Integrating over $d^3\vec x$ one finds the
 probability of having a monopole at $x=0$, while another is anywhere;
  fixing $x_4$ means that the probability refers to a given moment.
  Note the condition on magnetic charge $\int\tilde j_4 d^3x=1$,
   which yields the correct normalization.)
    Hence one gets an order of
 magnitude relation
  \be
   \frac{\nu(0)}{V_4}\sim \int d^3x<\tilde
 j(\vec x, x_4)\tilde j(0,x_4)>\sim D(0) T_g
  \label{7.4}
   \ee
    where we
 have assumed for $D(x)$ the form $D(x)=D(0)f(\frac{x}{T_g})$ and
 $f(y)$ is an exponential or Gaussian with $f(0)=1$.

 Finally, taking into account that $D(0)\sim g^2<F(0)F(0)>$ one
 obtains
 \be
 <\bar \psi\psi>\approx -\frac{g^2}{4\pi}<F^2(0)>T_g
 \label{7.5}
 \ee
 This estimate coincides with our result obtained from the
 quasiclassical calculation in [5]. Numerically
 (36) is -- $(300 MeV)^3$, i.e. a reasonable order of
 magnitude. Thus the very existence of the "wrong" correlator $D(x)$,
 violating the Abelian Bianchi identity may bring about monopole
 currents and ascociated with those zero modes.

   \section{Conclusion}

   The main question posed at the beginning of the paper --what is
the explicit form of dynamics, i.e. equations for a light quark  in
the field of a heavy static source. The answer to this question is
given by equations (15),(16).

 Those are Dirac--type equations with nonlocal nonlinear
 interaction. We have shown that these equations allow for
 solutions which yield scalar confining interaction. Therefore
 at large (as compared to $T_g$) distances one deals
 effectively with a local Dirac equation with a scalar
 potential. This is in contrast to the QED case considered in
 [1], where the one--body limit yields the well known Dirac
 equation with vector Coulomb potential. The local scalar
 potential was assumed for the heavy--light $q\bar q$ system
 in [3], where numerical and analytic results for spectra and
 wave functions are presented.

 The appearance of the scalar interaction provides (together
 with nonzero chiral condensate) an evidence for CSB, and the
 fact that these effects are proportional to the  string
 tension, tells about interconnection of CSB and confinement.
 There are several direction for the extension of this study.
 First, one should consider the case of 2 and 3 light quarks
 and derive the corresponding chiral Lagrangian. Second, with
 CSB at hand one can attack the problem of constituent quark
 mass. These topics are planned for subsequent publications.

 A large part of results of this paper have been obtained when
 the author was a guest of the Institute of Theoretical
 Physics in Utrecht. The cordial hospitality  and interesting
 discussions with Professor John Tjon  are gratefully
 acknowledged.

 This work was supported partially by the grants INTAS 93-79
 and RFFI grant 97-02-16404.


\begin{thebibliography}{99}
\bibitem{1}Yu.A.Simonov and J.A.Tjon, Ann Phys. (NY) {\bf 228} (1993)
1
\bibitem{2} R.P.Feynman, Phys. Rev. {\bf 80} (1950) 440; \\
J.Schwinger, Phys. Rev. {\bf 82 } (1951) 664
\bibitem{3}
V.D.Mur, V.S.Popov, Yu.A.Simonov and V.P.Yurov, JETP {\bf 78} (1994)
 1
\bibitem{4}
 Yu.A.Simonov, Physics - Uspekhi   {\bf 39} (1996) 313-336
\bibitem{5}
 Yu.A.Simonov, Theory of light quarks in the confining vacuum; hep-ph 9704301
 \bibitem{6}
N.G.Van Kampen, Phys. Rep. {\bf C24} (1976) 171; Physica {\bf 74}
(1974) 215\\
 Yu.A.Simonov , Yad. Fiz. {\bf 48} (1988) 1381; {\bf 50} (1989)
 213
 \bibitem{7}
H.G.Dosch, Yu.A.Simonov,  Phys. Lett. {\bf B 205} (1988) 339
 \bibitem{8}
 Y.Nambu and G.Iona--Lasinio, Phys. Rev. {\bf 122} (1961) 345\\
 S.P.Klevansky, Rev. Mod. Phys. {\bf 64} (1992) 649
 \bibitem{9}
 N.Brambilla and A.Vairo Phys. Lett. {\bf B 407} (1997) 167
  \bibitem{10}
  Yu.S.Kalashnikova and A.Nefediev, Potential regime for heavy quark dynamics
and  Lorentz nature of confinement, Hep-ph/9707490, Phys.Lett (in press)
     \bibitem{11} T.Banks and A.Casher, Nucl.  Phys.  {\bf B169} (1980) 102
  \bibitem{12}
  D.I.Diakonov and V.Yu. Petrov, {\bf 272} (1986) 457\\
E.Shuryak, A.Schaefer, hep-ph/9610245
\bibitem{13}
  A.Gonzalez--Arroyo, Yu.A.Simonov, Nucl. Phys. {\bf
 B 460} (1996) 429
 \bibitem{14}
  G.'tHooft, Nucl. Phys. {\bf B190} (1981) 455\\
  A.S.Kronfeld, G.Schierholz, U.Wiese, Nucl. Phys. {\bf B293} (1987)
  461
           \bibitem{15}
Yu.A.Simonov, Sov. J. Nucl. Phys. {\bf 42} (1985) 352
\end{thebibliography}
\end{document}